\documentclass[]{spie}  %>>> use for US letter paper
\usepackage[]{graphicx}
\usepackage{html}
\usepackage{hyperref}

\title{``Advanced'' data reduction for the AMBER
  instrument\thanks{illustrated by data acquired at the ESO Paranal
    observatory under the run number 074.A-9025.}} 

\author{Florentin Millour\supit{a}, Bruno Valat\supit{b}, Romain
  G. Petrov\supit{b}, Martin Vannier\supit{b}
  \skiplinehalf
  \supit{a}Max-Planck Institut for Radioastronomy, Auf dem h\"ugel, 69,
  53121, Bonn, Germany; \\
  \supit{b}Fizeau Laboratory, Nice university, Parc Valrose, Nice,
  France.
}

\authorinfo{Further author information: (Send correspondance to
  F. Millour)\\F. Millour: E-mail: fmillour@mpifr-bonn.mpg.de,
  Telephone: 49 228 525 188}

\begin{document} 
\maketitle

% {\bf Copyright 2008 Society of Photo-Optical Instrumentation
%  Engineers.} This paper will be published in the proceeding of SPIE
%  ``astronomical Telescopes and Instrumentation: Optical and Infrared
%  Interferometry'' and is made available as an electronic reprint
%  (preprint) with permission of SPIE. One print or electronic copy may
%  be made for personal use only. Systematic or multiple reproduction,
%  distribution to multiple locations via electronic or other means,
%  duplication of any material in this paper for a fee or for commercial
%  purposes, or modification of the content of the paper are
%  prohibited.

%%%%%%%%%%%%%%%%%%%%%%%%%%%%%%%%%%%%%%%%%%%%%%%%%%%%%%%%%%%%% 
\begin{abstract}
  The amdlib AMBER data reduction software is meant to produce AMBER
  data products from the raw data files that are sent to the PIs of
  different proposals or that can be found in the ESO data
  archive. The way defined by ESO to calibrate the data is
  to calibrate one science data file with a calibration one,
  observed as close in time as possible. Therefore, this scheme does
  not take into account instrumental drifts, atmospheric variations or
  visibility-loss corrections, in the current AMBER data processing
  software, amdlib.

  In this article, we present our approach to complement this default
  calibration scheme, to perform the final steps of data reduction,
  and to produce fully calibrated AMBER data products. These
  additional steps include: an overnight view of the data structure
  and data quality, the production of night transfer functions from
  the calibration stars observed during the night, the correction of
  additional effects not taken into account in the standard AMBER data
  reduction software such as the so-called "jitter" effect and the
  visibility spectral coherence loss, and finally, the production of
  fully calibrated data products.

  All these new features are beeing implemented in the modular
  pipeline script \texttt{amdlibPipeline}, written to complement the
  amdlib software.
\end{abstract}

% >>>> Include a list of keywords after the abstract 


%%%%%%%%%%%%%%%%%%%%%%%%%%%%%%%%%%%%%%%%%%%%%%%%%%%%%%%%%%%%% 

\section{INTRODUCTION}
\label{sect:intro}% \label{} allows reference to this section

The AMBER data reduction software:
\texttt{amdlib}\cite{2007A&A...464...29T}, features a new type of
algorithm, the so-called P2VM
algorithm\cite{2004SPIE.5491.1222M}, which performs a direct fit of
the observed fringes using a template fringe pattern recorded during
the instrument calibration and produces a series of observables;
namely, the squared visibility, the closure phase, and the differential
phase\cite{2006-ITHD_FMillour}. The standard calibration plan handles the
AMBER signal from the raw data to raw averaged visibilities OI DATA
files. Though a simple calibration script
(\texttt{amdlibDivideOiData}) is included in the \texttt{amdlib}
software, a complete calibration procedure is not yet provided.

In this document, we present our approach to produce calibrated OI
data files. It consists of several recipes, just like for the standard
\texttt{amdlib} software, to organize, evaluate, plot,
and finally, calibrate the data. This philosophy allows one to switch easily
from one recipe to another if necessary.
This leads to the modular structure of the example pipeline script
\texttt{amdlibPipeline}\footnote{currently available at
  \url{http://www.mpifr-bonn.mpg.de/staff/fmillour/}}, written in the
scientific language \texttt{yorick}, to perform the final steps of data
reduction after an \texttt{amdlib} data reduction session.

This paper is organized into three different sections: 
\begin{itemize}
\item A section about file management and coping with different types
  of calibrations (different DITs, different modes, use of BCD),
\item The rise of advanced calibration methods: ``coherence length''
  and ``jitter'' effects correction,
\item How to cope with calibration stars and transfer function, as
  well as a template for data calibration.
\end{itemize}

\section{DATA MANAGEMENT}

The AMBER user generally faces a first difficulty when receiving the
ESO DVDs: one gets a series of files sorted by night or by observing
block number, with a name reflecting the observing time:
{``AMBER.yyyy-mm-ddThh:mm:ss.sss.fits''}. The \texttt{amdlib}
software, or even the ESO tool \texttt{gasgano}, already present some
tools to ease the sorting and use of the AMBER raw data files:
colouring the different files types, or ``automatic'' data reduction
scripts. Here we propose to add some more comfort to the user to
evaluate information on the overall observing night. Therefore, this
section is meant to answer the following questions: How many stars and
calibrators were observed during a night ? What is the UV coverage for
the science stars ? Was the observing night a success or not ?

\subsection{Find a star in a night directory}

The small script \texttt{amdlibListStars} rapidly gives the
information of which stars were observed during a night and whether
they are tagged as ``SCIENCE'' or ``CALIB''. It makes use of the log
file produced by the standard \texttt{amdlib } software and just
compiles it into a short stars list. An example of the type of result
one can get from this script is given below:

\begin{verbatim}
> amdlibListStars
 Choose the product directory (OI_DATA)
    
 The file/directory chosen is:
    /disk/DATA/2007-05-16/
 HD124454 CALIB
 Alpha_Arae SCIENCE
 HD164371 CALIB
 hd-test 
 HD-93813 
\end{verbatim}

Here one can see that the actually observed stars are HD 124454,
$\alpha$ Arae, and HD 164371. The star $\alpha$ Arae is the science
star, whereas HD 124454 and HD 164371 are two calibrators. The other
two targets: hd-test and HD 93813 were meant to be observed, but the
observations were not executed (no \texttt{CALIB} or \texttt{SCIENCE}
tag associated).

In the same philosophy, the script \texttt{amdlibListAllStars} does
the same, but on a series of night data directories, and produces a
r\'esum\'e file \texttt{starList.txt} listing the stars and giving
the different directory where we find them.

\subsection{Check file tags and fix them}

For some old data, or during a technical night where all the stars are tagged
as ``CALIB'', it may be convenient to tag again the corresponding
files to ``SCIENCE''. The script called
\texttt{amdlibConvertScienceToCalib} is made for that purpose and
applies new tags to all the files linked to a given star. It makes use
of the precedent function and fits files read/write routines.

\subsection{Sorting files by integration time, instrument mode, etc.}

This is often convenient to sort the data files with specific
properties such as the instrument modes, whether the fringe tracker
was used or not, etc. The function \texttt{sortAMBERFiles} allows one to
sort the files by creating sub-directories and producing symbolic
links to the original files. By default, it will sort the files by star
name, but it is often convenient to sort the files by exposure
time instead (as the transfer functions at different exposure times do not
match). In table \ref{tab:ESOkeys}, one can find the relevant fits
keywords for sorting the files, with a short description of the
sorting criteria. For example, if one wants to sort files by
exposure time, the following command should be typed in yorick: 

\begin{verbatim}
    > sortAMBERFiles, keyword="ESO DET DIT", keyDec=1
\end{verbatim}

In case one works with compressed files (the suffix of the files
is then \texttt{.fits.gz} instead of \texttt{.fits}), the symbolic
links will not work with the amdlib software. Alternatively, one can move
the files instead of creating symbolic links using the option
\texttt{symbolic=0}.

%% This table is carefully placed in the source file to make 
%% it appear at bottom of page, but above the footnotes.  
%% Use of [h] in following command forces table to appear "here".
\begin{table}[htbp]
  \caption{Typical ESO FITS keywords usable for sorting the AMBER files.} 
  \label{tab:ESOkeys}
  \begin{center}       
    \begin{tabular}{|l|l|} %% this creates two columns
      %% |l|l| to left justify each column entry
      %% |c|c| to center each column entry
      %% use of \rule[]{}{} below opens up each row
      \hline
      FITS keyword & Short description \\
      \hline
      ESO OBS TARG NAME & Star name\\
      \hline
      ESO DET DIT & Detector Integration Time (exposure time)\\
      \hline
      DATE-OBS & Date of observation\\
      \hline
      ESO INS MODE & Instrument mode: whether 2 or 3 telescopes, spectral
      resolution, etc.\\
      \hline 
      ESO DEL FT STATUS & FINITO (fringe tracker) status: ON or OFF.\\
      \hline 
      ESO INS OPTI7 NAME & BCD (Beam Commuting Device) status: IN or OUT\\
      \hline 
    \end{tabular}
  \end{center}
\end{table} 

\subsection{UV coverage during a night}

\begin{figure}
  \begin{center}
    \begin{tabular}{c}
      \includegraphics[height=10cm]{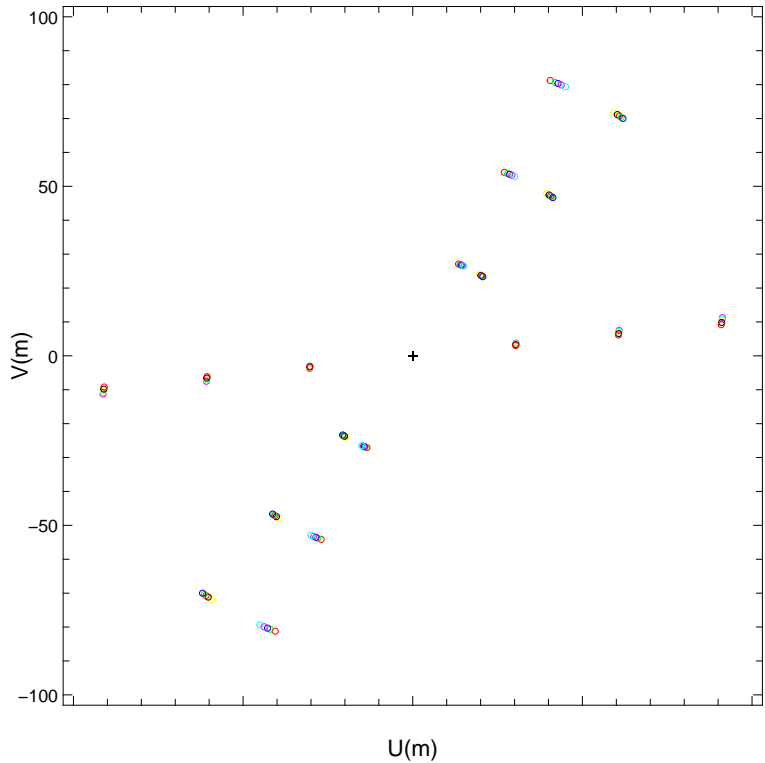}
    \end{tabular}
  \end{center}
  \caption[The UV plane coverage on a star during a full night] 
  % >>>> use \label inside caption to get Fig. number with \ref{}
  { \label{fig:UVcov} 
    UV coverage of a star observed in 2007 in a full night. Here, 3
    consecutive shots were recorded with the large aligned triplet.
  }
\end{figure}

An interesting feature is to know what type of UV coverage was
observed for a given star in an observing night. The small script
\texttt{amdlibPlotUV\_Night} allows one to plot the UV coverage for
all the science stars present in a data directory (see
Fig.~\ref{fig:UVcov}). One option also allows one to plot the
calibration stars UV coverage.

\section{ADVANCED DATA PROCESSING}

The amdlib software lacks two important steps in the data processing
scheme; namely, the coherence loss correction and the so-called
``jitter'' correction. We present here the developments we made to
correct these two effects and the implications they have on the data
processing.

\subsection{Correcting the visibilities from coherence length losses}

This step is fundamental for low spectral resolution data, as the
coherence length $L_c$ is smaller than the maximum atmospheric
optical path difference (OPD) stroke, with typical values of 30$\mu$m
and 35-40$\mu$m, respectively. As shown in Millour et
al.\cite{2007arXiv0705.1636M}, the probable shape of the spectral
channels of AMBER is a Gaussian curve of FWHM $L_c$. A first idea is
to correct instantaneous squared visibilities from this visibility
loss using the frame-by-frame OPD measurement.

\begin{equation}
  \rho(\delta) = exp \left[ - 2 \ln(2) \left(\frac{\delta}{
        L_c}\right)^2 \right]
  \label{eq:vAtt}
\end{equation}

However, the very important thing to notice here is that the OPD
value, $\delta$, given in equation \ref{eq:vAtt} is an \emph{absolute}
value of relative to the white fringe. Instead, the OPD measurement in
AMBER is always made relative to the P2VM reference. Therefore, this
P2VM OPD effect has to be taken into account as it can strongly affect
the OPD measurements (see Fig.~\ref{fig:calibVis}). Moreover, since
the AMBER fibers are not exactly of equal length, a chromatic OPD is
induced in the fringe pattern. This leads to curved fringes and, thus,
an OPD measurement dependent on $\lambda$. Therefore, one needs a
$\lambda$-dependent OPD measurement on the P2VM to correct for it.

\begin{figure}[htbp]
  \centering
  \begin{tabular}{cc}
    \includegraphics[height=0.48\hsize,origin=c, angle=-90]{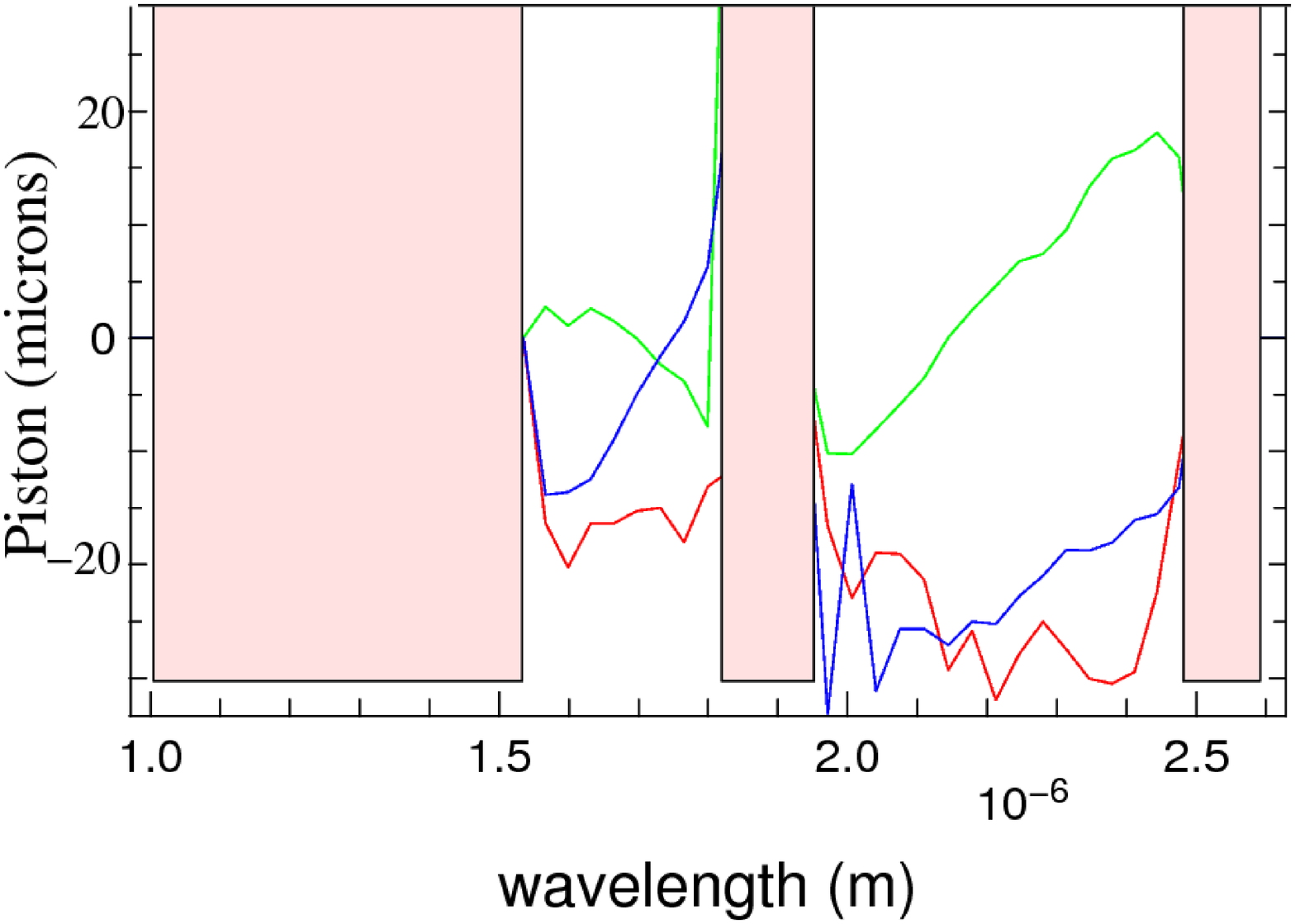} &
    \includegraphics[width=0.48\hsize, height=0.34\hsize,origin=c, angle=0]{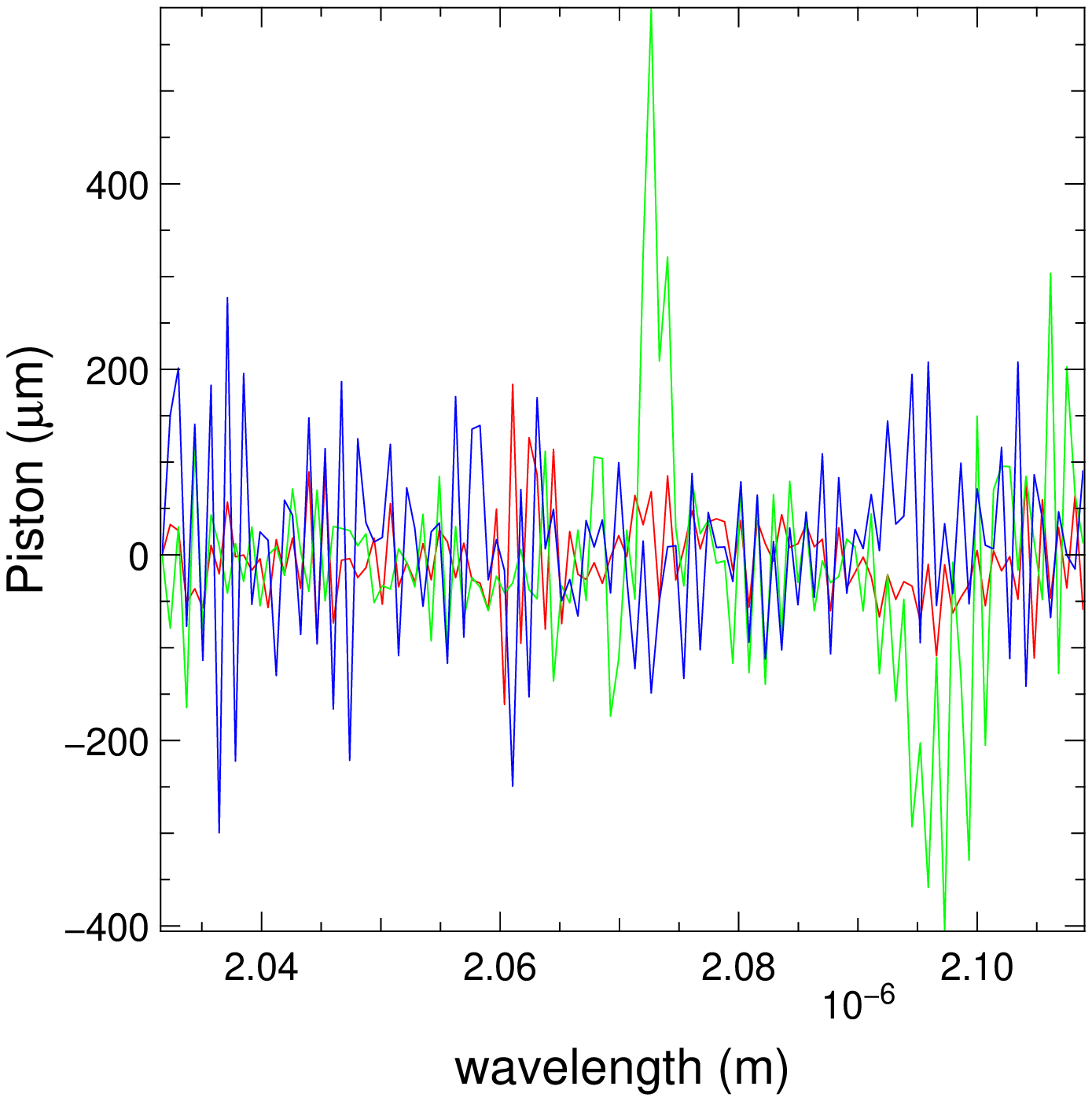}
  \end{tabular}
  \caption{
    \emph{Left}: Example of a non-zero OPD on a P2VM. The masked zones
    correspond to the inter-bands and the J band where the OPD
    cannot be computed.
    \emph{Right}: Example of a $\gamma^2$ Vel observation, where the
    object's phase can introduce chromatic ``OPD'' as high as
    500$\mu$m, which is no longer negligible for this medium
    resolution data (same as the one of Fig. \ref{fig:calibVis}).
  }
  \label{fig:chromPist}
\end{figure}

For the same reason, because the atmosphere also induces chromatic
effects, one needs to take into account the chromatic OPD on the
target itself to correct for the curved fringes in the white fringe
position estimation. Figure \ref{fig:chromPist} illustrates these two
problems for two extreme cases:
\begin{itemize}
\item for a low resolution P2VM, where the chromatic P2VM
  OPD can be as high as 20$\mu$m due to a non-optimal OPD alignment
  of the AMBER instrument,
\item for a medium resolution data-set, where the OPD
  effect coming from the object's phase itself (peaks in the green
  base) can produce an off-centering of the white light fringe
  compared to the usual OPD computation.
\end{itemize}

The function \texttt{amdlibCorrectCoherenceLength} is a proposition
for such an implementation: it uses a \texttt{RAW} OI data file,
containing instantaneous squared visibilities and OPD measurements
as input, and produces a raw OI data file called \texttt{COR}, where the
visibilities have been corrected according to equation \ref{eq:vAtt},
using an OPD measurement both on the associated P2VM and on the curved
fringes information contained in the differential phase of the
object. Then, the usual steps of \texttt{amdlib} (frame selection and
data averaging) can be proceed on these files.

\subsection{Correcting the visibilities from the ``jitter'' effect}
\label{sect:jitter}  

Another origin of visibility loss is the fringe motion during one
single exposure time, usually called ``jitter'' because it reflects
the fringe smearing onto the instrument detector during one frame
integration.

Based on a proposition from Malbet et al.\cite{1998ApJ...507L.149M},
we developed an estimator based on the assumption that the jitter
effect comes only from a regular phase shift between two consecutive
frames. To do so, we compute a cross-product of the raw complex
coherent flux between two consecutive frames:

\begin{equation}
  W^{12}_{ij} = C_{ij}(\lambda, t_1) \times C_{ij}^{*}(\lambda, t_2)
  \label{eq:interspec}
\end{equation}

\begin{figure}
  \begin{center}
    \begin{tabular}{ccc}
      \includegraphics[height=5.5cm]{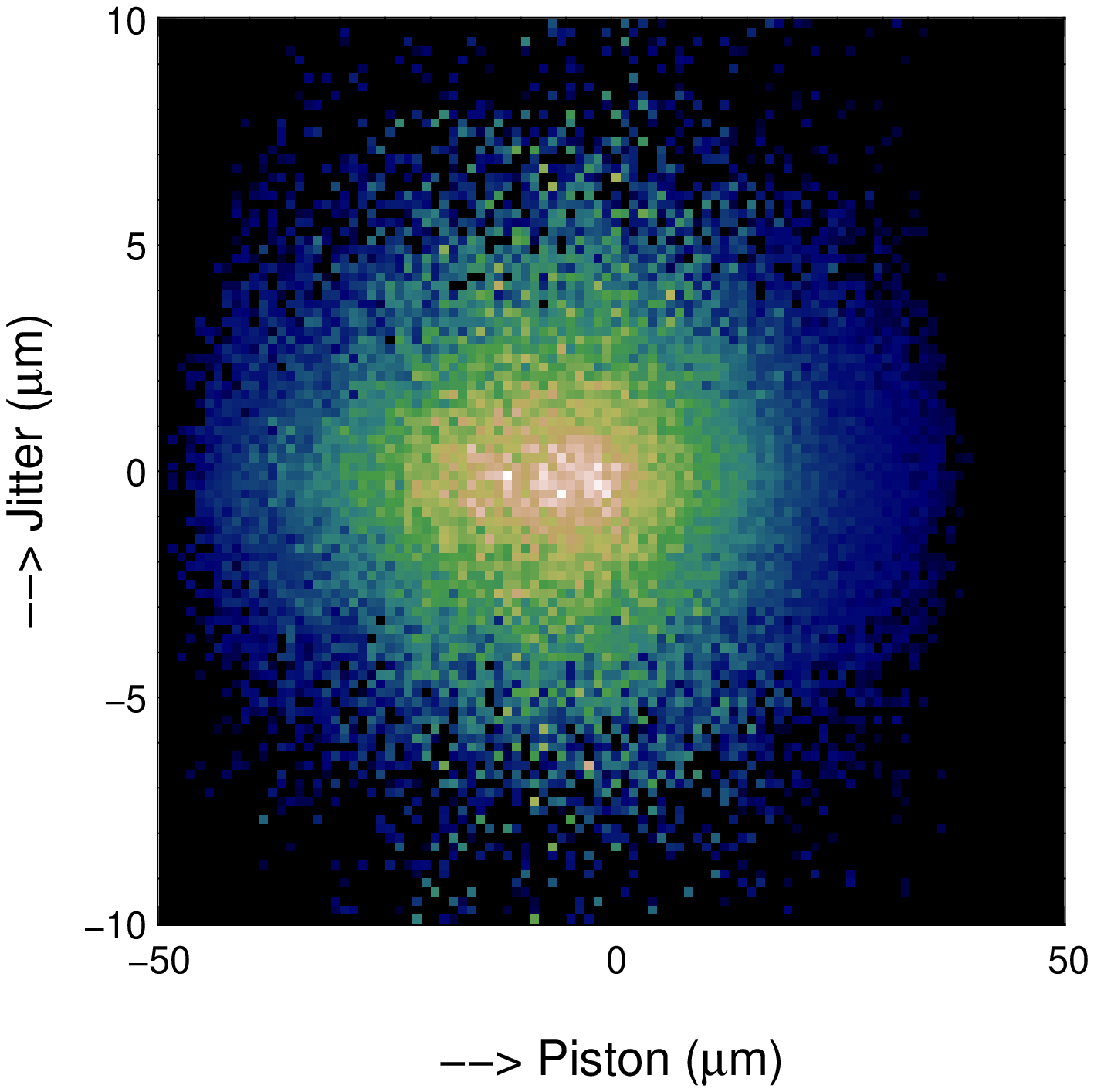}&
      \multicolumn{2}{c}{\includegraphics[height=5.5cm]{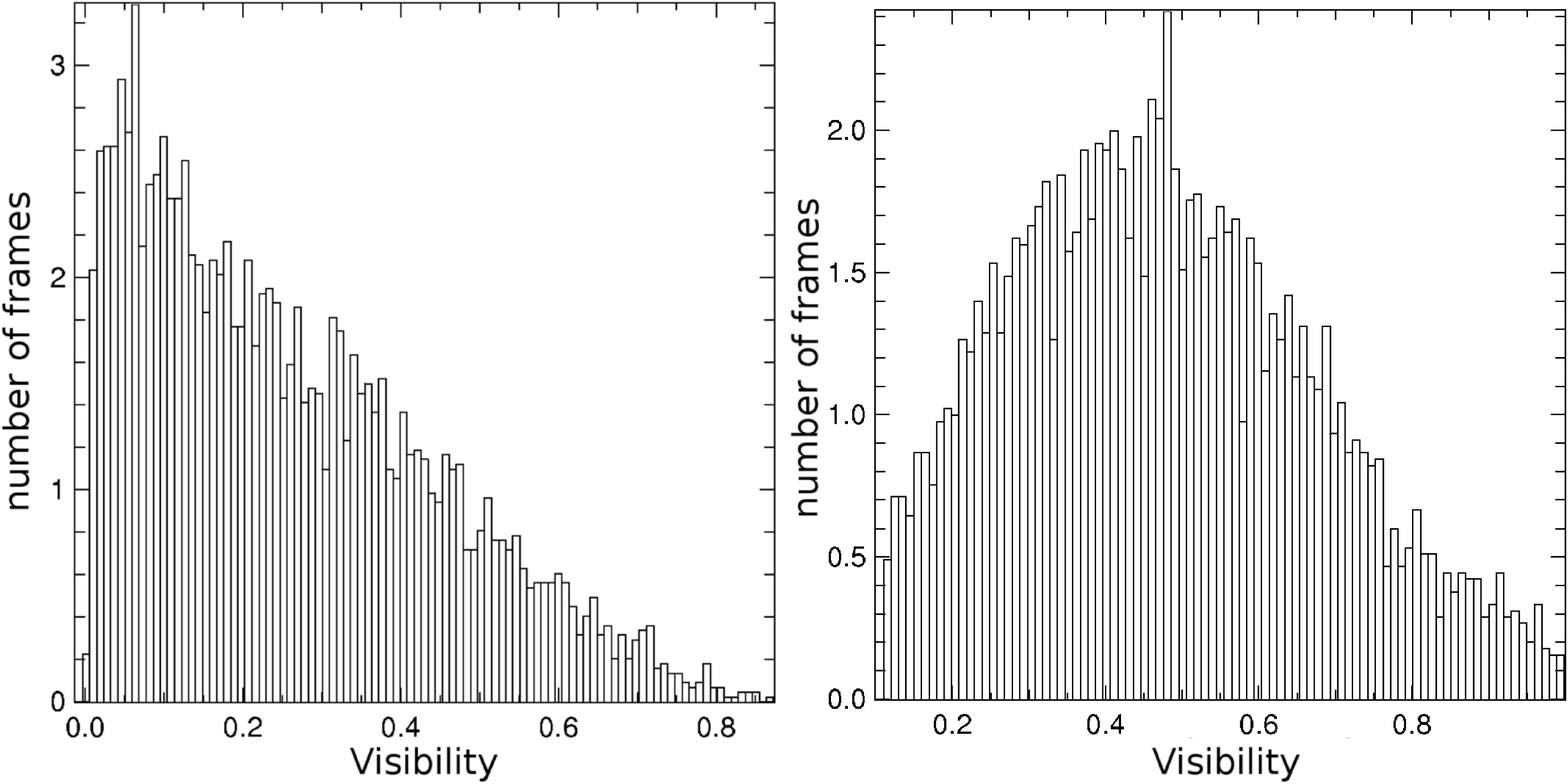}}\\
      \includegraphics[height=5.5cm]{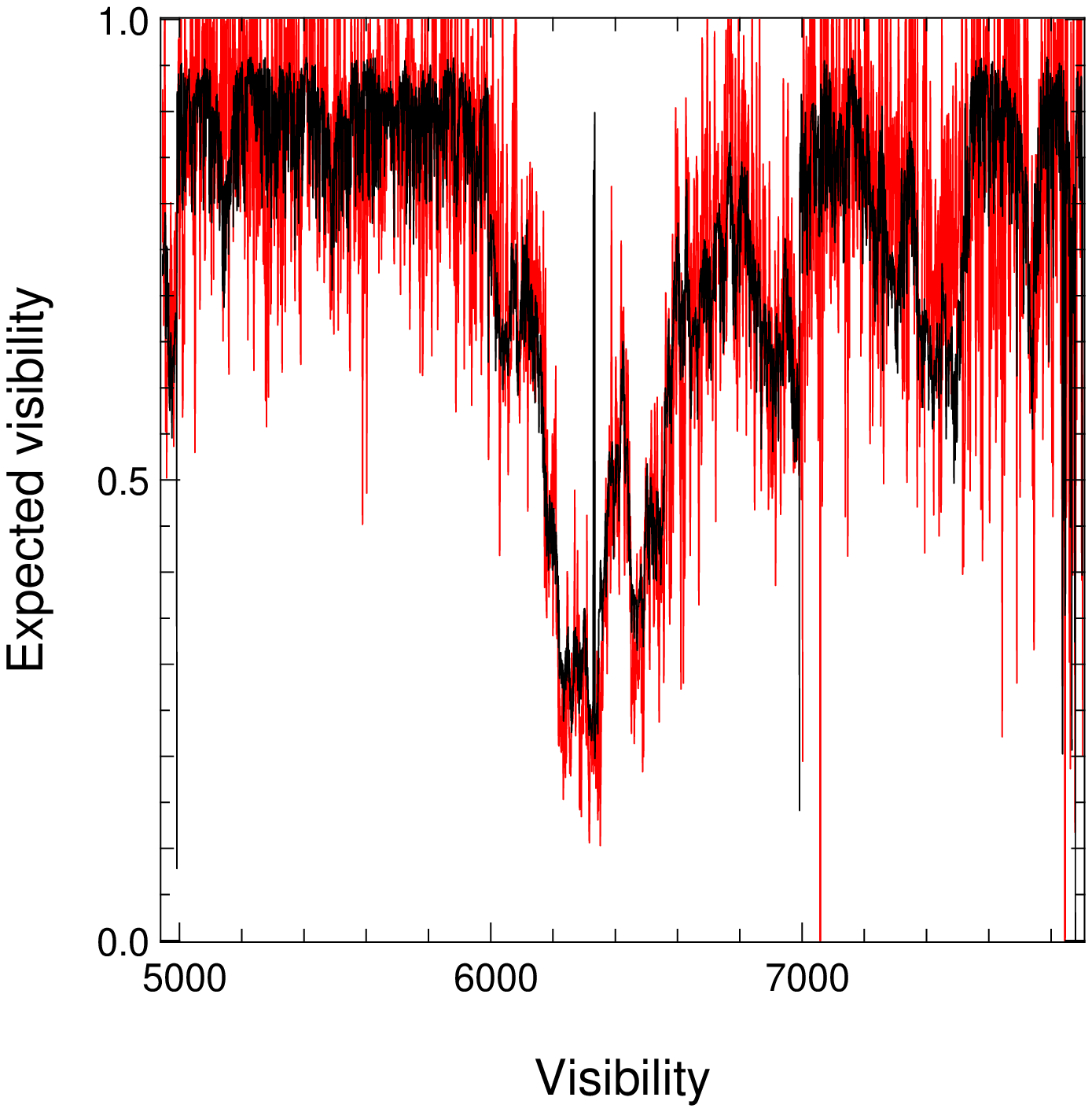}&
      \includegraphics[height=5.5cm]{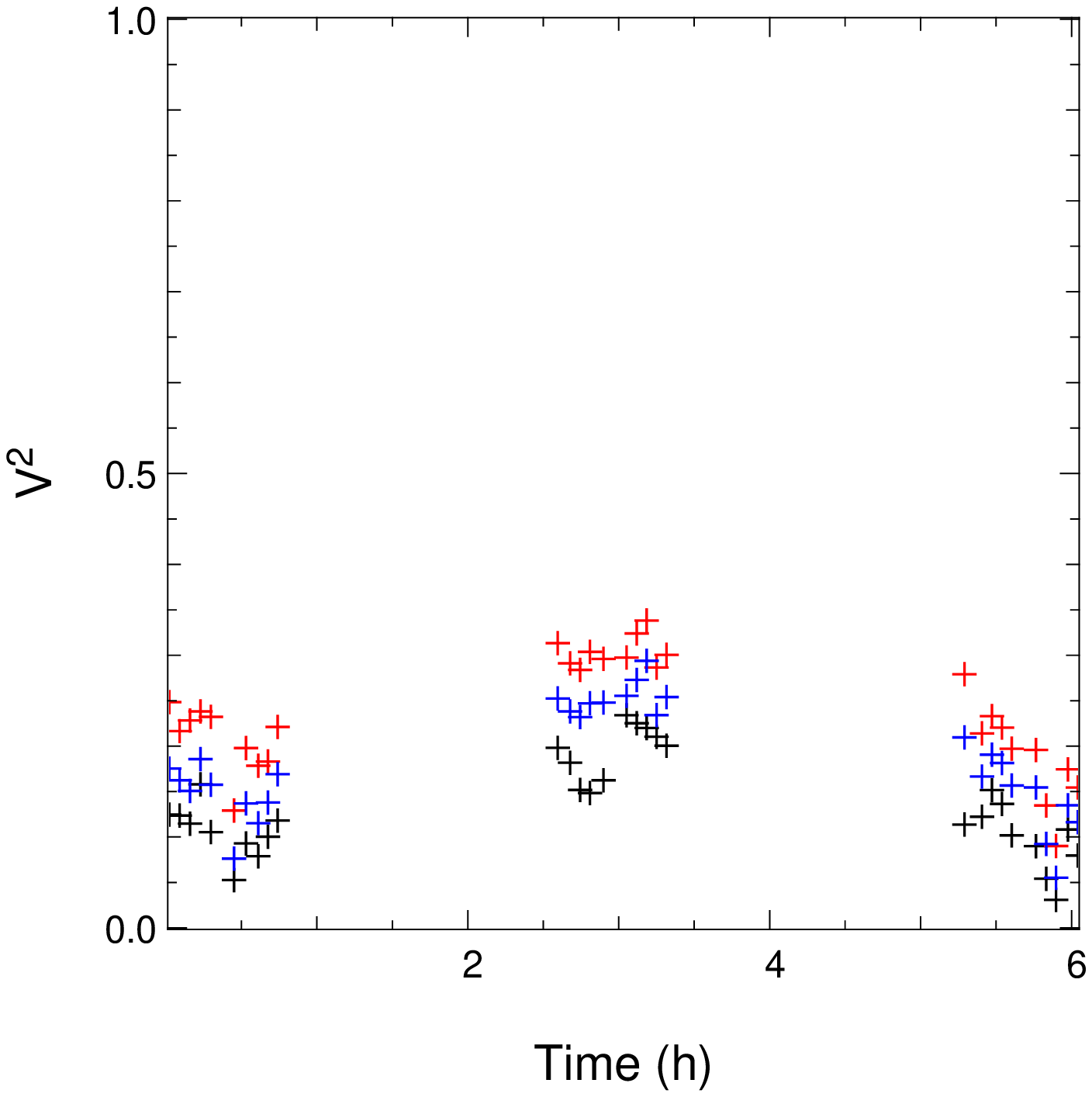}&
      \includegraphics[height=5.5cm]{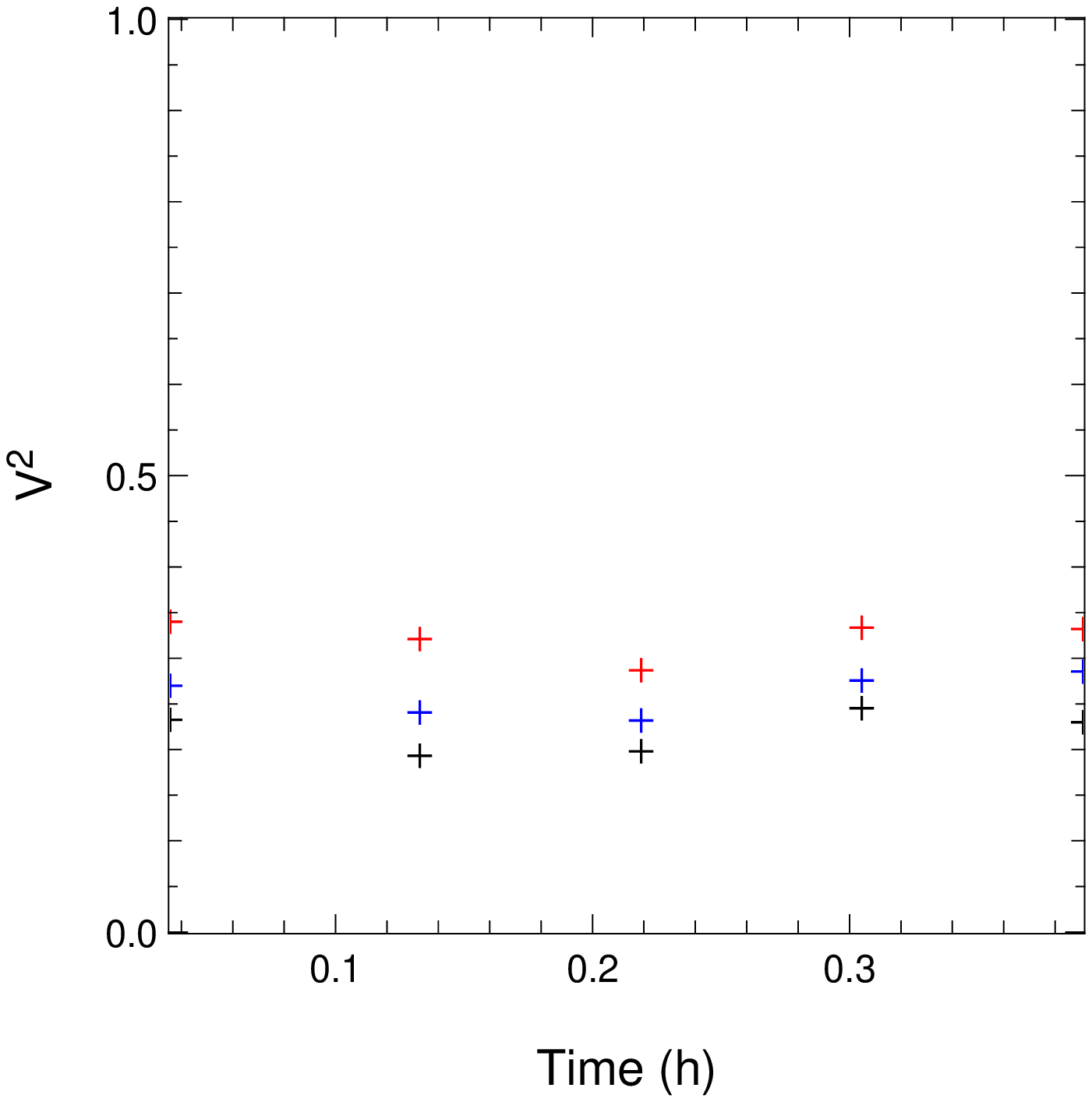}
    \end{tabular}
  \end{center}
  \caption{\emph{Top-left:} Correlation plot showing the squared
    visibility as a function of our jitter estimator (vertical
    direction) and OPD (horizontal direction). One can see that in
    both directions, the squared visibility can be empirically
    described by a 2-D Gaussian-like surface.
    \emph{Top-right:} Squared visibility histograms without (left) and with
    (right) jitter and OPD loss correction. The histogram switches from
    a log-normal distribution to a Gaussian distribution, which is much
    easier to reduce than before.
    \emph{Bottom-left:} a time-sequence of squared visibilities (in
    red) compared to the expected visibility loss from jitter and
    piston variations.
    \emph{Bottom-middle:} example of a 10hrs time-sequence for
    uncorrected $V^2$ (standard amdlib output without any frame
    selection, black crosses), with the coherence length corrected (in
    blue) and with the coherence length + jitter corrected (in red,
    upper curve). The typical scatter of visibilities is divided by 2
    in this case (30\% to 15\% relative error).
    \emph{Bottom-right:} a shorter time-sequence on very good quality
    data, showing a much higher improvement: from 16\% relative error
    to less than 4\%.
  }
  \label{fig:B.V.Transfert_function}
\end{figure}

The phase of this cross-product gives an estimate of the phase
difference between these two frames. An OPD is then computed on this
cross-product, which is the difference of OPD between the two
frames. Then, a correlation plot between the OPD, OPD difference, and
squared visibility is built (see Fig.~\ref{fig:B.V.Transfert_function}). In the
OPD direction, the correlation is ``well known'' and can be computed
according to equation \ref{eq:vAtt}. In the ``jitter'' direction, we
find that the squared visibility decrease can be empirically described
with a Gaussian-like shape. Therefore, one can describe the squared
visibility as a function of the coherence loss and of the jitter as a
2-D Gaussian curve. Therefore, one can use this
now known dependence to correct the visibilities for jitter and
coherence losses, and compare it with uncorrected data. This is shown in
Fig.~\ref{fig:B.V.Transfert_function} on the right, where the original
squared visibility histogram, which has a log-normal distribution,
becomes a Gaussian histogram after correction. This example test has
been repeated on many data sets and always shows the same improvement
of the visibilities histogram.

Moreover, we find that the transfer function (i.e. the squared
visibility vs time, for one star) is more stable after this jitter
correction than if only the coherence length correction is applied to
the data. Therefore, we can claim here that we have found a way of
estimating and correcting the jitter effect, which was a burden for
low spectral resolution AMBER data until now. The next step for now is
to implement this method in a robust way into the standard AMBER data
processing software.

\section{CALIBRATING THE DATA}

Data calibration can be done in several ways, using classical optical
interferometry calibration techniques (like in
Perrin 2003\cite{2003A&A...400.1173P}) or speckle data calibration. The aim
of this section is not to explain these different calibration
techniques, but to propose a data handling framework where all types
of calibrations could be done more easily.

\subsection{Finding/registering the diameter of calibration stars}

This step consists in browsing all the calibration stars observed
during a night and finding their associated angular diameters to
correct in a further step their observed visibility from the expected
one. Usually, it is left to the user to do such step, but it is not as
easy as one might think. In several cases (browsing archival data,
etc.), one does not have direct access to the calibration stars
diameters. Moreover, many catalog databases (reported in
table~\ref{tab:CalCat}) and calibrator-seeking software ({\tt
  ASPRO}\footnote{\url{http://www.mariotti.fr/aspro\_page.htm}}
(based on the work of Bonneau et al.\cite{Bonneau2006}), {\tt
  CalVin}\footnote{\url{http://www.eso.org/observing/etc/}} or {\tt
  getCal}\footnote{\url{http://msc.caltech.edu/software/getCal/}})
exist, which make the whole calibration diameter finding step
complicated.

\begin{table}[htbp]
  \caption{Available catalogs of stars giving access to apparent
    stellar diameters. ind. stands for indirect estimation method,
    LBSI stands for Long Baseline Stellar Interferometry and Lun. occ
    stands for Lunar Occultation.}
  \label{tab:CalCat}
  \begin{center}       
    \begin{tabular}{|l|c|c|c|c|c|c|c|c|} %% this creates two columns
      %% |l|l| to left justify each column entry
      %% |c|c| to center each column entry
      %% use of \rule[]{}{} below opens up each row
      \hline
      Name     & Nr of stars & Techniques used & Vizier name & Reference\\
      \hline
      ESO VLTI & 191        & LBSI            & J/A+A/434/1201 &
      Richichi et al., 2005, A\&A, 434, 1201\\
      \hline
      Bord\'e    & 374        & ind.            & J/A+A/393/183 & Bord\'e
      et al., 2002, A\&A, 393, 183\\
      \hline
      M\'erand   & 948        & ind.            & J/A+A/433/1155 &
      M\'erand et al., 2005, A\&A, 433, 1155\\
      \hline
      CHARM    & 3248       & LBSI, Lun. occ. & J/A+A/386/492 &
      Richichi et al., 2002, A\&A, 386, 492\\
      \hline
      CHARM2   & 8231       & LBSI, Lun. occ., ind. & J/A+A/431/773 &
      Richichi et al., 2005, A\&A, 431, 773\\
      \hline
      CADARS   & 13570      & LBSI, ind.      & II/224/cadars & Pasinetti et
      al. 2001, A\&A 367, 521\\
      \hline
    \end{tabular}
  \end{center}
\end{table} 

The approach used here is to produce a local database containing
sufficient information, taken from the fits file header and used to
identify the stars. This information is namely the star name used
during the observation and its coordinates. This allows one to query
the different catalogs listed in table \ref{tab:CalCat} from CDS and
fill-in missing information in the database, used afterwards to
correct the calibrator visibilities and produce a transfer function.

In our pipeline, this database is a text file which can be
automatically produced with the function \texttt{amdlibGetCalibzSize},
given that an Internet connection is available. This script perform a
series of queries on CDS in the available catalogs of interferometric
calibrators using the vizquery
\footnote{\url{http://vizier.u-strasbg.fr/doc/vizquery.htx}} utility
from CDS.

\subsection{Computing a transfer function}

This step aims at building the measured instrumental response to
calibrate afterwards the science data. The function
\texttt{amdlibComputeTransFunc} uses the precedent calibrators
database to identify the stars from the fits keywords contained in the
files' headers. The transfer function is simply computed accounting
for a uniform disk (UD) shape for the stellar disk. As most of the
catalogs give UD diameters, this is relevant for the further
steps. As many catalogs are used, it happens sometimes that the
diameter was measured several times. In this case, we take the average
diameter and compare the diameter dispersion with the existing
errors. The expected calibration stars visibility is then computed
according to the following equation:

\begin{equation}
  V^2 = \left| 2 \frac{J_1\left(\frac{\pi B
          \theta}{\lambda}\right)}{\frac{\pi B \theta}{\lambda}} \right|^2
  \label{eq:visDiam}
\end{equation}

The averaged measured visibility of the star is divided by this
expected visibility and the result is saved in a new OI FITS
file. This allows us to keep this calibration step made
for further analysis using any of the existing calibration method. The
variance on visibility from the diameter error, if present, is also
computed and added to the visibility variance, to take this
calibration error into account in the transfer function.

\subsection{Checking the transfer function}

Checking the quality of the transfer function may be a difficult task
since AMBER is a spectroscopic instrument. In principle, one would have
to plot 2D-maps giving the transfer function at each wavelength for
each time of the night. Because this may not be easy to plot, we
deliberately propose two functions to plot the transfer function
projected onto the time axis or onto the wavelength axis:

\begin{itemize}
\item \texttt{amdlibPlotTransFuncTime} plots wavelength-averaged
  visibilities as a function of time, to check the overnight stability
  of the instrumental+atmospheric transfer function. As default,
  the entire wavelength range available is averaged for all of the
  files, but one can choose a wavelength interval with the
  \texttt{wlenRange} option to get only the wavelengths of interest.
\item \texttt{amdlibPlotTransFuncWlen} plots a projection of the
  transfer function on the wavelength axis. This plot is similar to
  the one which is often used for the MIDI instrument, except that
  here we get 3 baselines and a closure phase.
\end{itemize}

\begin{figure}[htbp]
  \centering
  \begin{tabular}{cc}
    \includegraphics[width=0.48\hsize]{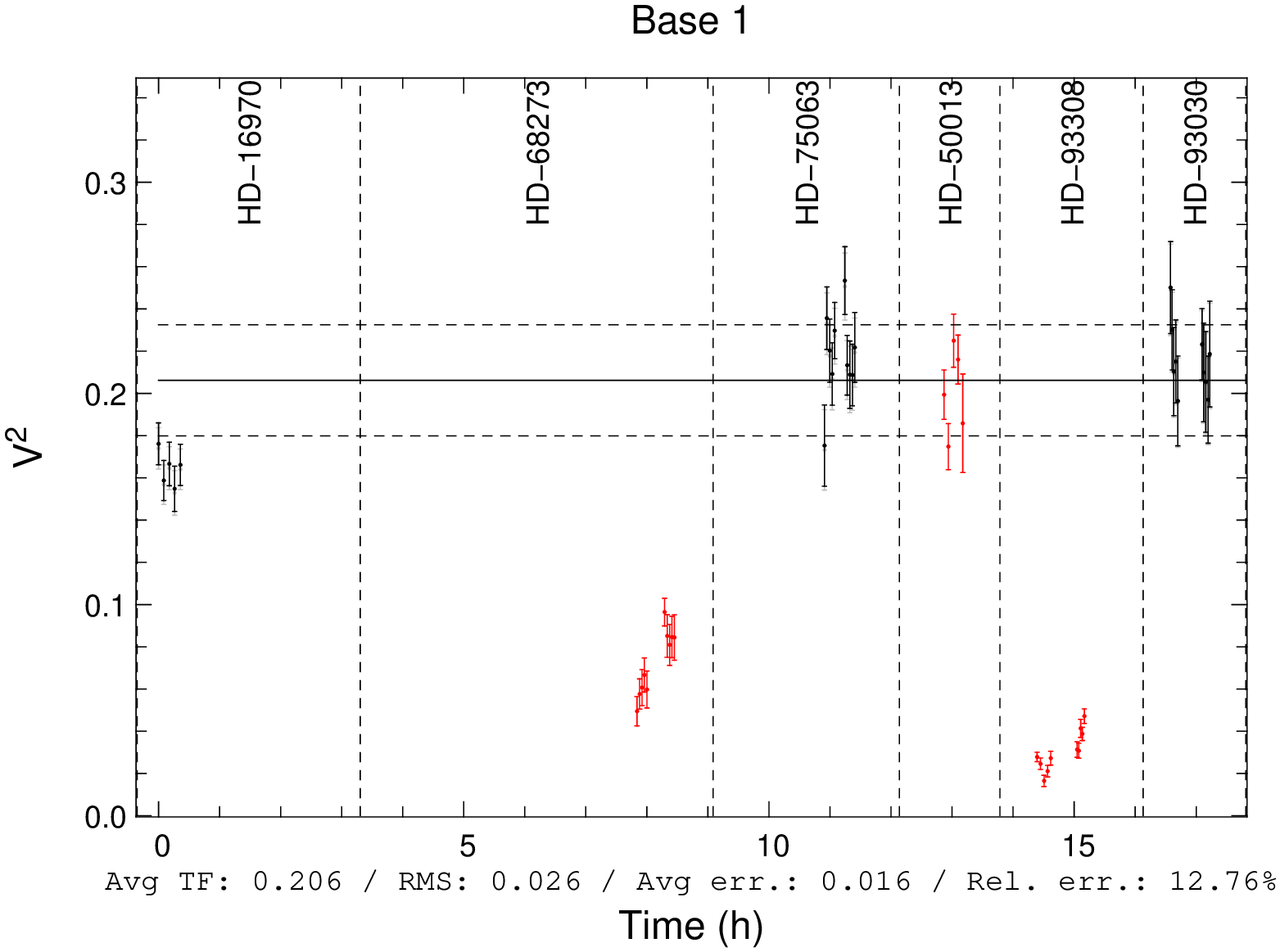} &
    \includegraphics[width=0.48\hsize]{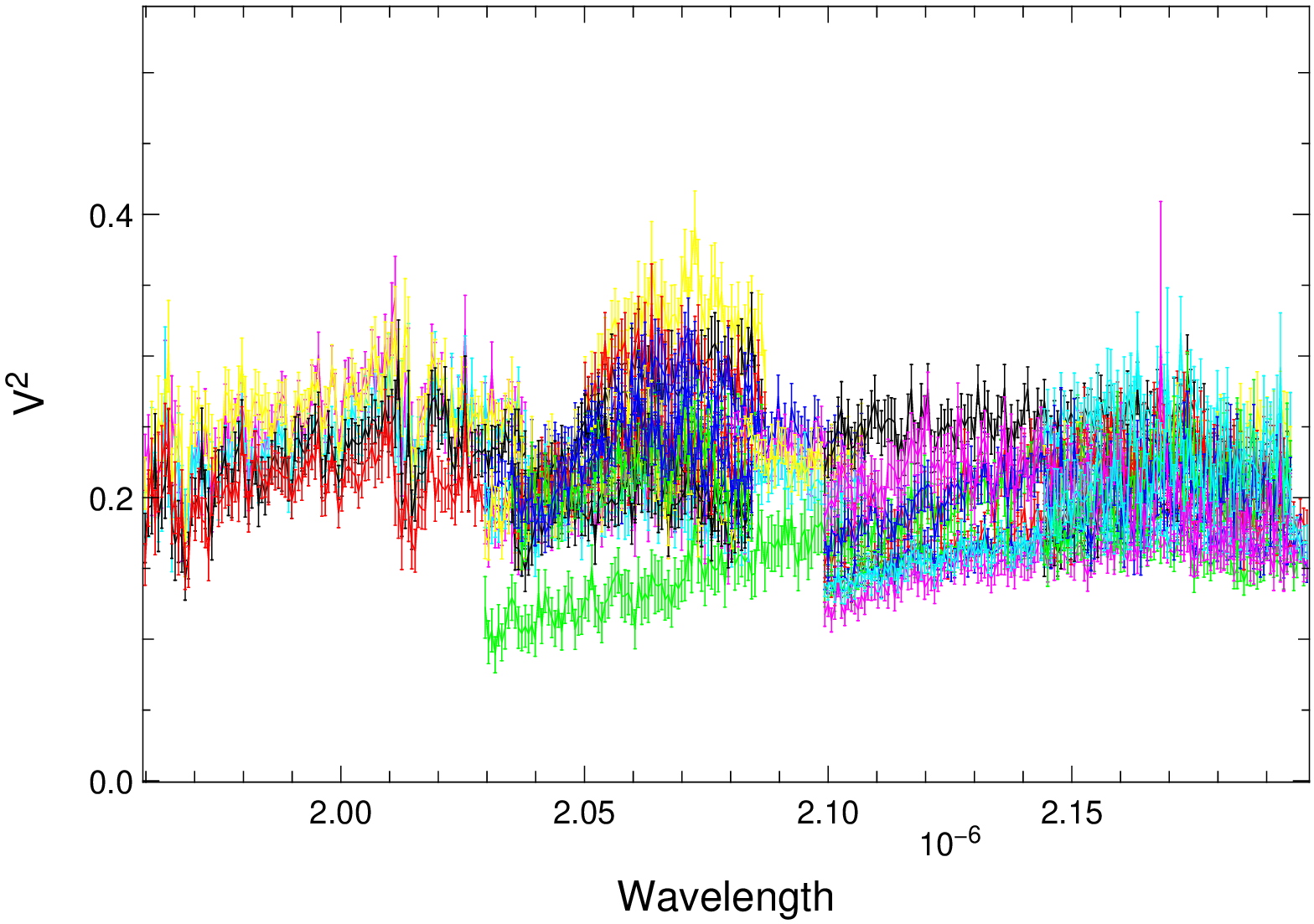}
  \end{tabular}
  \caption{
    Left: The transfer function of 25/12/2004 plotted by
    \texttt{amdlibPlotTransFuncTime}, in black, showing the stability
    over the night, to be compared with the science data published in
    several articles\cite{2007A&A...464...73M, 2007A&A...464..107M,
      2007A&A...464...87W} in red. The star HD16970 was observed at
    low spectral resolution and pollutes the whole night transfer
    function.
    Right: The same transfer function plotted with
    \texttt{amdlibPlotTransFuncWlen}. Here the science data is not
    plotted as it would be more complicated to disentangle from the
    transfer function.
  }
  \label{fig:transFunc}
\end{figure}

One can see the result in figure \ref{fig:transFunc} for the
first science night of AMBER (25/12/2004), showing that the
process works fine, even for old data sets.

\subsection{Calibrating the data}

The last step of the interferometric data reduction is the calibration
of the science data itself. As explained in
Perrin 2006\cite{2003A&A...400.1173P}, the calibration of visibilities needs an
accurate time-resolved estimate of the instrument and atmosphere
transfer function. This is not the case for the AMBER instrument, as
the minimum time between a science star and a calibration star is
30\,mn (due to operation strategies and overheads in the VLTI).% ,
% whereas, for example, the typical significant Seeing change in Paranal
% is about 10\,mn (Kishimoto, private communication). Therefore, one
% cannot directly apply this method for calibrating the AMBER data.

Here, we present a simple approach for the AMBER data calibration,
that can be used as an example for More sophisticated calibration
algorithms. The function \texttt{amdlibCalibrateOiDataFMi} calibrates
the data using the very simple assumption that the transfer function 
is constant throughout the night and that the variability appears as a
scattering of visibilities around its average. This assumption is
thought to be relevant in the current VLTI case, where large time
intervals (more than 10mn) are used between calibration stars,
preventing the interpolation of the transfer function with respect to
time.

\begin{figure}[htbp]
  \centering
  \begin{tabular}{ccc}
    \includegraphics[width=0.3\textwidth]{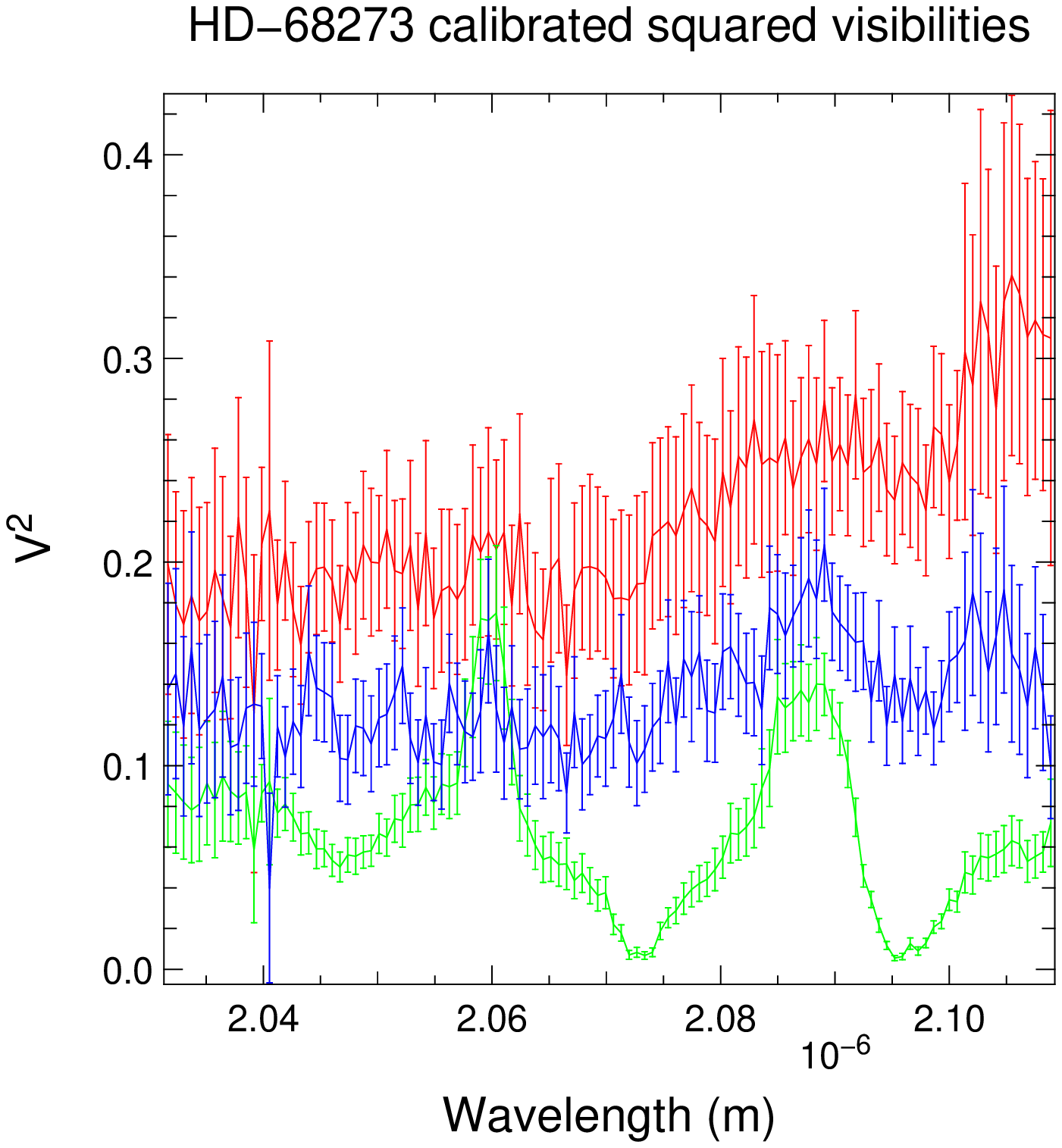} &
    \includegraphics[width=0.3\textwidth]{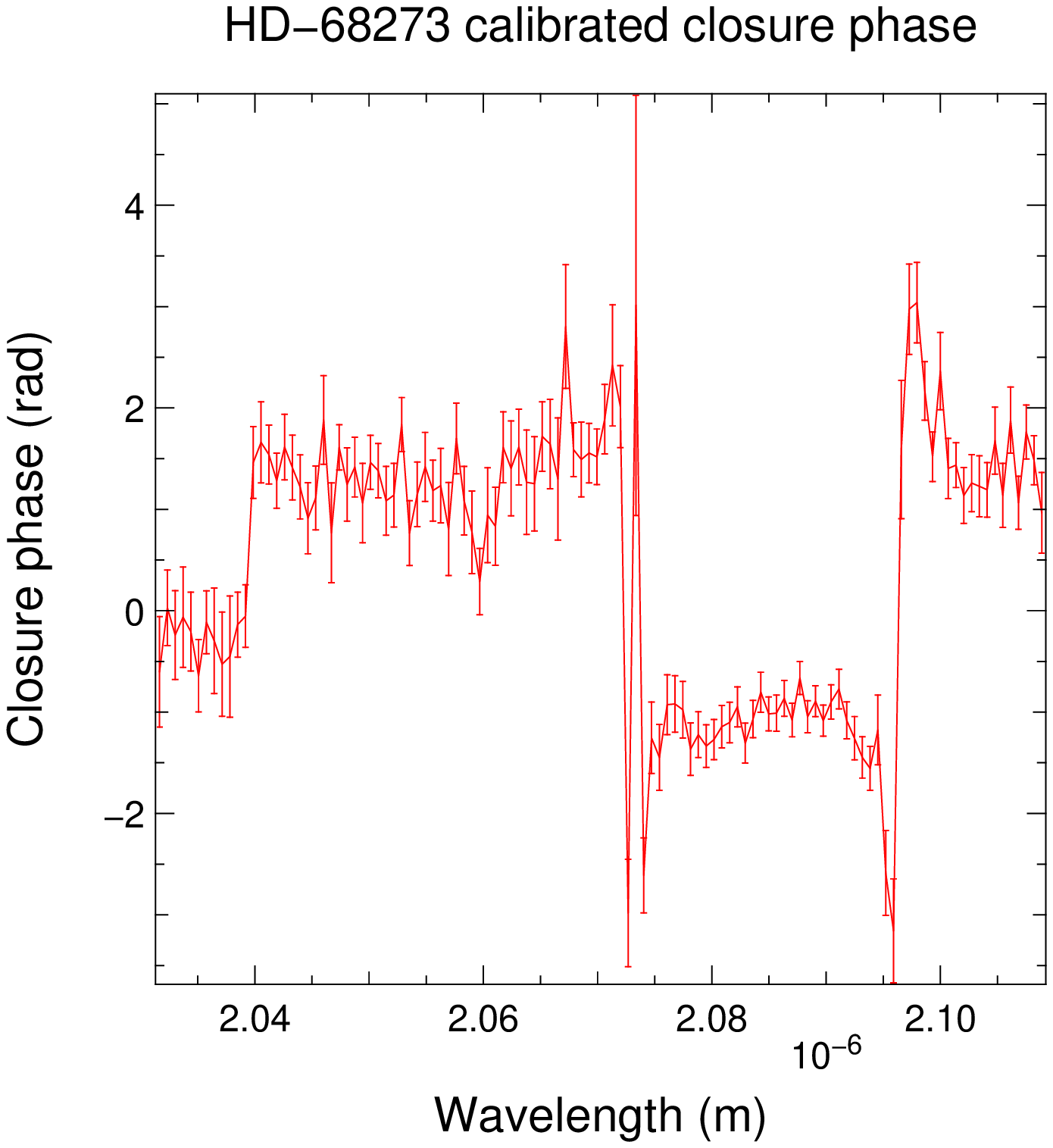} &
    \includegraphics[width=0.3\textwidth]{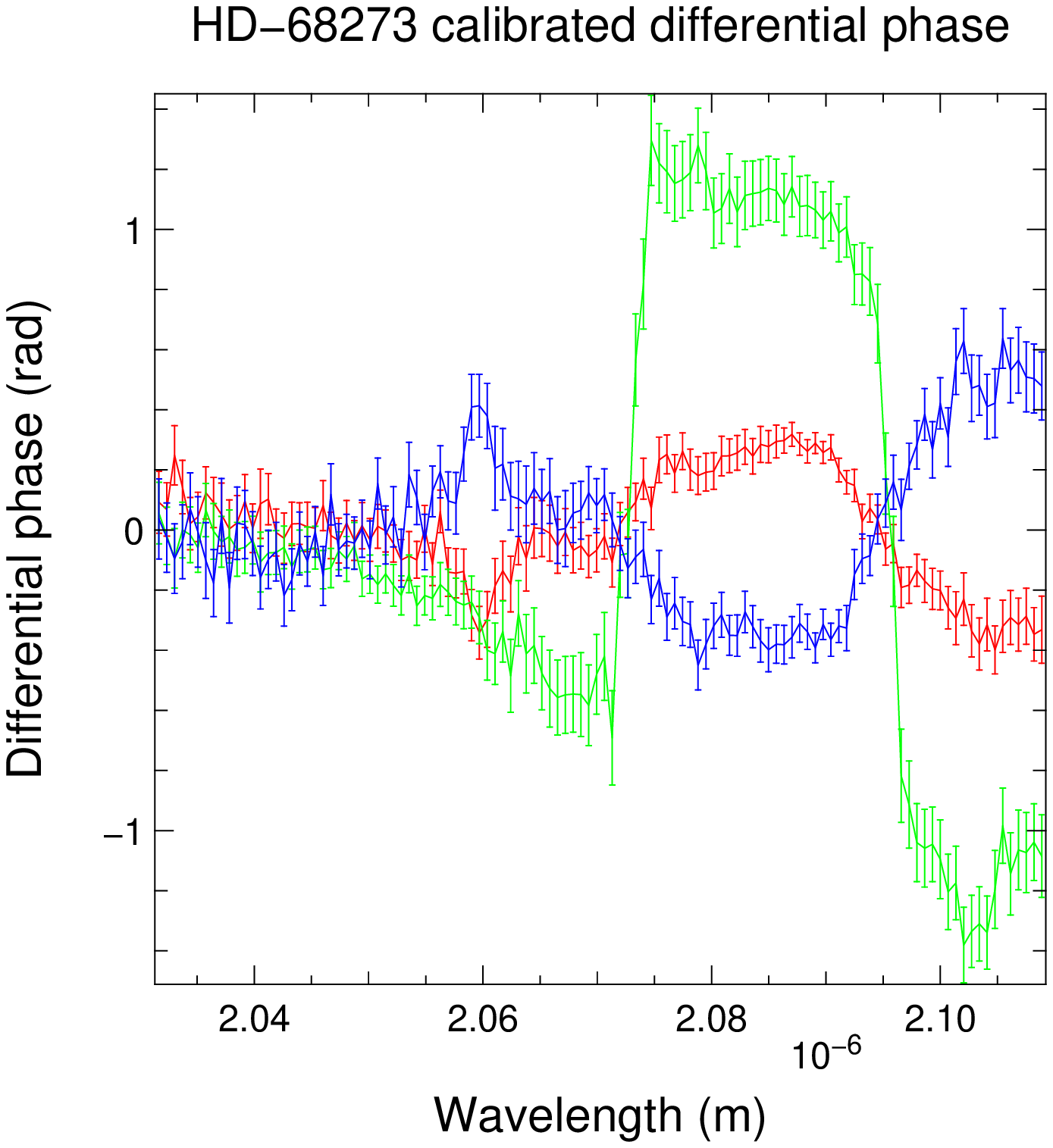}
  \end{tabular}
  \caption{
    Left: Calibrated squared visibilities
    Middle: Calibrated closure phase
    Right: Calibrated differential phases
  }
  \label{fig:calibVis}
\end{figure}

Therefore, this script loads the transfer function and the
science-tagged files, computes an average transfer function and a
scatter for each science data wavelength, and divides the latter by
the first. This function provides calibrated OI fits files as a result
that can be used for science. An example of the resulting calibrated
observables plot is shown in figure \ref{fig:calibVis} and can be
compared with Millour et al.\cite{2007A&A...464..107M}.

\section{Conclusion}

The standard default calibration scheme proposed for AMBER
(calibrating one file with another) is obviously insufficient to get
realistic error bars, and an overnight calibration quality estimate.

We presented a proposition for a user-friendly pipeline framework
for the amdlib software. We are able, with the proposed scripts and
an Internet connection, to produce science-grade data with realistic
error bars in a semi-automated way, as it is already done for the
first data reduction steps in \texttt{amdlib}.

This article is thought to be a starting point for further reflexions
about AMBER data calibration. Further work could allow one to perform
the AMBER data reduction in a fully automated way, allowing in the
future the production of archived data products, in the same way as it
is done with the ESO imaging and spectroscopic instruments (WFI,
FEROS, etc.).

%%%%%%%%%%%%%%%%%%%%%%%%%%%%%%%%%%%%%%%%%%%%%%%%%%%%%%%%%%%%% 
%%%%% References %%%%%

\bibliographystyle{spiebib}   %>>>> makes bibtex use spiebib.bst
\bibliography{biblio}   %>>>> bibliography data in report.bib

\end{document}